\documentstyle[12pt]{article}

\font\twlgot =eufm10 scaled \magstep1 \font\egtgot =eufm8
\font\sevgot =eufm7 \font\twlmsb =msbm10 scaled \magstep1
\font\egtmsb =msbm8 \font\sevmsb =msbm7

\newfam\gotfam
\def\pgot{\fam\gotfam\twlgot}
\textfont\gotfam\twlgot \scriptfont\gotfam\egtgot
\scriptscriptfont\gotfam\sevgot
\def\got{\protect\pgot}
\newfam\msbfam
\textfont\msbfam\twlmsb \scriptfont\msbfam\egtmsb
\scriptscriptfont\msbfam\sevmsb
\def\Bbb{\protect\pBbb}
\def\pBbb{\relax\ifmmode\expandafter\Bb\else\typeout{You cann't use
Bbb in text mode}\fi}
\def\Bb #1{{\fam\msbfam\relax#1}}

\newcommand{\gd}{{\got d}}
\newcommand{\gP}{{\got P}}

\newcommand{\gL}{{\got L}}

\def\thebibliography#1{\bigskip\section*{}\bigskip\list
{$^{\arabic{enumi}}$}{\settowidth\labelwidth{#1}\leftmargin\labelwidth
\advance\leftmargin\labelsep
\usecounter{enumi}}
\def\newblock{\hskip .11em plus .33em minus .07em}
\sloppy\clubpenalty4000\widowpenalty4000 \sfcode`\.=1000\relax}

\def\op#1{\mathop{\fam0 #1}\limits}

\newcommand{\id}{{\rm Id\,}}

\newcommand{\im}{{\rm Im\,}}
\newcommand{\nm}[1]{|{#1}|}
\newcommand{\beq}{\begin{equation}}
\newcommand{\eeq}{\end{equation}}
\newcommand{\ben}{\begin{eqnarray}}
\newcommand{\een}{\end{eqnarray}}
\newcommand{\be}{\begin{eqnarray*}}
\newcommand{\ee}{\end{eqnarray*}}
\newcommand{\bea}{\begin{eqalph}}
\newcommand{\eea}{\end{eqalph}}
\newcommand{\cA}{{\cal A}}

\newcommand{\cP}{{\cal P}}

\newcommand{\cL}{{\cal L}}

\newcommand{\cE}{{\cal E}}

\newcommand{\cQ}{{\cal Q}}
\newcommand{\cF}{{\cal F}}
\newcommand{\cS}{{\cal S}}
\newcommand{\cC}{{\cal C}}
\newcommand{\cO}{{\cal O}}

\newcommand{\cG}{{\cal G}}

\newcommand{\cN}{{\cal N}}
\newcommand{\bL}{{\bf L}}

\newcommand{\bE}{{\bf E}}

\newcommand{\al}{\alpha}
\newcommand{\vr}{\varrho}
\newcommand{\bt}{\beta}
\newcommand{\dl}{\delta}
\newcommand{\la}{\lambda}
\newcommand{\La}{\Lambda}
\newcommand{\f}{\phi}

\newcommand{\m}{\mu}

\newcommand{\g}{\gamma}

\newcommand{\th}{\theta}

\newcommand{\vt}{\vartheta}

\newcommand{\up}{\upsilon}
\newcommand{\lng}{\langle}
\newcommand{\rng}{\rangle}

\newcommand{\si}{\sigma}
\newcommand{\Si}{\Sigma}
\newcommand{\w}{\wedge}

\newcommand{\wh}{\widehat}
\newcommand{\ol}{\overline}
\newcommand{\dr}{\partial}
\newcommand{\ar}{\op\longrightarrow}
\newcommand{\llr}{\op\longleftarrow}
\newcommand{\lto}{\leftarrow}
\newcommand{\ot}{\otimes}

\newcommand{\rdr}{\stackrel{\leftarrow}{\dr}{}}

\let\ssection=\section
\renewcommand{\section}{\setcounter{equation}{0}\ssection}

\newcounter{eqalph}
\newcounter{equationa}
\newcounter{remark}
\newcounter{example}
\newcounter{theorem}
\newcounter{proposition}
\newcounter{lemma}
\newcounter{corollary}
\newcounter{definition}
\newenvironment{eqalph}{\stepcounter{equation}
\setcounter{equationa}{\value{equation}} \setcounter{equation}{0}

\begin{eqnarray}}{\end{eqnarray}\setcounter{equation}{\value{equationa}}}

\def\theremark{\arabic{remark}}

\def\thetheorem{\arabic{theorem}}

\newenvironment{proof}{
{\it Proof:}}{}

\newenvironment{theo}{\refstepcounter{theorem}
{\bf Theorem \thetheorem:}}{}

\newcommand{\mar}[1]{}

\begin{document}
\hbox{}

{\parindent=0pt

{\large\bf Lagrangian BV quantization and Ward identities}
\bigskip

{\sc D. Bashkirov}\footnote{Electronic mail: bashkir@phys.msu.ru},
{\sc G. Sardanashvily}\footnote{Electronic mail:
sard@grav.phys.msu.su}

{\sl Department of Theoretical Physics, Moscow State University,
117234 Moscow, Russia}

\bigskip
\bigskip

The Ward identities are the relations which the complete Green
functions of quantum fields satisfy if an original classical
Lagrangian system is degenerate. A generic degenerate Lagrangian
system of even and odd fields is considered. It is characterized
by a hierarchy of reducible Noether identities and gauge
supersymmetries parameterized by antifields and ghosts,
respectively. In the framework of the BV quantization procedure,
an original degenerate Lagrangian is extended to ghosts and
antifields in order to satisfy the master equation. Replacing
antifields with gauge fixing terms, one comes to a non-degenerate
Lagrangian which is quantized in the framework of perturbed QFT.
This Lagrangian possesses a BRST symmetry. The corresponding Ward
identities are obtained. They generalize Ward identities in the
Yang--Mills gauge theory to a general case of reducible gauge
supersymmetries depending on derivatives of fields of any order. A
supersymmetric Yang-Mills model is considered.

 }

\bigskip
\bigskip

\noindent {\bf I. INTRODUCTION}
\bigskip

In a general setting, by Ward identities are meant the relations
which the complete Green functions of quantum fields satisfy if an
original classical Lagrangian system is degenerate.

A generic degenerate Lagrangian system of even and odd fields is
considered \cite{cmp05,jmp05,jmp05a}. Its Euler--Lagrange operator
satisfies nontrivial Noether identities. They need not be
independent, but obey the first-stage Noether identities, which in
turn are subject to the second-stage ones, and so on. Being
finitely generated, the Noether and higher-stage Noether
identities are parameterized by the modules of antifields
(Section III). The Noether second theorem states the relation
between these Noether identities and the reducible gauge
supersymmetries of a degenerate Lagrangian system parameterized by
ghosts (Section IV).

In the framework of the BV quantization of a degenerate Lagrangian
system \cite{bat,gom,fust}, its original Lagrangian $L$ is
extended to the above mentioned ghosts and antifields in order to
satisfy the so-called classical master equation (Section V).
Replacing antifields with gauge fixed terms, one comes to a
non-degenerate gauge-fixing Lagrangian $L_{GF}$ (\ref{z16}) which
can be quantized in the framework of perturbed QFT (Section VII).
This Lagrangian possesses the gauge-fixing BRST symmetry $\wh u$
(\ref{z17}). We obtain the corresponding Ward identities
(\ref{z61}) -- (\ref{z62}) for complete Green functions of quantum
fields (Section VIII). They generalize the Ward identities in the
Yang--Mills gauge theory \cite{slavn,becc}, whose gauge symmetries
are irreducible, independent of field derivatives, and they form a
finite-dimensional Lie algebra.

Note that the BRST transformation $\wh u$ is necessary non-linear
and, moreover, it need not maintain the measure term in a
generating functional of Green functions. Therefore, the Ward
identities (\ref{z61}) -- (\ref{z62}) contain an anomaly in
general.  Another anomaly can issue from the fact that, after
regularization, a Lagrangian $L_{GF}$ need not be BRST invariant.
Here, we however do not concern a regularization procedure.

In Section IX, an example of supersymmetric Yang-Mills model is
considered.

\bigskip
\bigskip

\noindent {\bf II. GRASSMANN-GRADED LAGRANGIAN SYSTEMS}
\bigskip

Bearing in mind the BV quantization, we consider a Lagrangian
field system on $X=\Bbb R^n$, $2\leq n$, coordinated by $(x^\la)$.
Such a Lagrangian system is algebraically described in terms of a
certain bigraded differential algebra (henceforth BGDA)
$\cP^*[Q;Y]$ \cite{cmp05,jmp05,jmp05a}. Unless otherwise stated,
by a gradation throughout is meant $\Bbb Z_2$-gradation.

Let $Y\to X$ be an affine bundle coordinated by $(x^\la,y^i)$
whose sections are even classical fields, and let $Q\to X$ be a
vector bundle coordinated by $(x^\la,q^a)$ whose sections are odd
classical fields. Let $J^rY\to X$ and $J^rQ\to X$, $r=1,\ldots,$
be the corresponding $r$-order jet bundles, endowed with the
adapted coordinates $(x^\la, y^i_\La)$ and $(x^\la, q^a_\La)$,
respectively, where $\La=(\la_1...\la_k)$, $|\La|=k$,
$k=1,\ldots,r$, are symmetric multi-indices. The index $r=0$
conventionally stands for $Y$ and $Q$. For each $r=0,\ldots,$ we
consider a graded manifold $(X,\cA_{J^rQ})$, whose body is $X$ and
the ring of graded functions consists of sections of the exterior
bundle
\be
\w (J^rQ)^*=\Bbb R\op\oplus_X (J^rQ)^*\op\oplus_X\op\w^2
(J^rQ)^*\op\oplus_X\cdots,
\ee
where $(J^rQ)^*$ is the dual of a vector bundle $J^rQ\to X$. The
global basis for $(X,\cA_{J^rQ})$ is $\{x^\la,c^a_\La)$,
$|\La|=0,\ldots,r$. Let us consider the graded commutative
$C^\infty(X)$-ring $\cP^0[Q;Y]$ generated by the even elements
$y^i_\La$ and the odd ones $c^a_\La$, $|\La|\geq 0$. The
collective symbols $s^A_\La$ further stand for these elements,
together with the symbol $[A]=[s^A_\La]$ for their Grassmann
parity. In fact, $\cP^0[Q;Y]$ is the $C^\infty(X)$-ring of
polynomials in the graded elements $s^a_\La$.

Let $\gd \cP^0[Q;Y]$ be the  Lie superalgebra of (left) graded
derivations of the $\Bbb R$-ring $\cP^0[Q;Y]$, i.e.,
\be
u(ff')=u(f)f'+(-1)^{[u][f]}fu(f'), \qquad f,f'\in \cP^0[Q;Y],
\qquad u\in \gd\cP^0[Q;Y].
\ee
Its elements take the form
\mar{w5}\beq
 u=u^\la\dr_\la + \op\sum_{0\leq|\La|} u_\La^A\dr^\La_A,
 \qquad u^\la, u_\La^A
 \in \cP^0[Q;Y], \label{w5}
\eeq
where $\dr^\La_A(s_\Si^B)=\dl_A^B\dl^\La_\Si$ up to permutations
of multi-indices $\La$ and $\Si$. By a summation over a
multi-index $\La$ is meant separate summation over each index
$\la_i$. For instance, we have the total derivatives
\be
d_\la =\dr_\la + \op\sum_{0\leq|\La|} s_{\la+\La}^A\dr^\La_A\in
\gd\cP^0[Q;Y],
\ee
where $\la+\La$ denotes the multi-index
$(\la,\la_1,\ldots,\la_k)$.

With the Lie superalgebra $\gd \cP^0[Q;Y]$, one can construct the
minimal Chevalley--Eilenberg differential calculus
\be
0\to \Bbb R\to \cP^0[Q;Y] \ar^d \cP^1[Q;Y]\ar^d\cdots
\cP^2[Q;Y]\ar^d\cdots
\ee
over the ring $\cP^0[Q;Y]$. It is a desired BGDA $\cP^*[Q;Y]$. Its
elements $\f\in \cP^k[Q;Y]$ are graded $\cP^0[Q;Y]$-linear
$k$-forms on $\gd \cP^0[Q;Y]$ with values in $\cP^0[Q;Y]$. The
graded exterior product $\w$ and the even Chevalley--Eilenberg
coboundary operator $d$, called the graded exterior differential,
obey the relations
\be
 \f\w\f' =(-1)^{|\f||\f'| +[\f][\f']}\f'\w
\f, \qquad  d(\f\w\f')= d\f\w\f' +(-1)^{|\f|}\f\w d\f',
\ee
where $|.|$ denotes the form degree. Since $\cP^*[Q;Y]$ is a
minimal differential calculus over $\cP^0[Q;Y]$, it is generated
by the elements $dx^\la$, $ds^A_\La$ dual of $\dr_\la$,
$\dr_A^\La$, i.e.,
\be
 \f= \op\sum \f_{A_1\ldots A_r\la_1\ldots\la_k}^{\La_1\ldots
\La_r} ds_{\La_1}^{A_1}\w\cdots\w ds_{\La_r}^{A_r}\w
dx^{\la_1}\w\cdots \w dx^{\la_k}, \qquad \f\in \cP^0[Q;Y].
\ee
In particular, the graded exterior differential takes the familiar
form
\be
 d\f=dx^\la\w \dr_\la\f +\op\sum_{0\leq|\La|}
ds^A_\La\w\dr_A^\La \f.
\ee
Let $\cO^*X$ be the graded differential algebra of exterior forms
on $X$. There is the natural monomorphism $\cO^*X\to \cP^*[Q;Y]$.

Given a graded derivation $u$ (\ref{w5}) of the $\Bbb R$-ring
$\cP^0[Q;Y]$, the interior product $u\rfloor\f$ and the Lie
derivative $\bL_u\f$, $\f\in\cP^*[Q;Y]$, are defined by the
formulas
\be
&& u\rfloor \f=u^\la\f_\la +
\op\sum_{0\leq|\La|}(-1)^{[\f^\La_A][A]}u^A_\La\f_A^\La, \qquad
\f\in \cP^1[Q;Y],\\
&& u\rfloor(\f\w\si)=(u\rfloor \f)\w\si
+(-1)^{|\f|+[\f][u]}\f\w(u\rfloor\si), \qquad \f,\si\in
\cP^*[Q;Y], \\
&& \bL_u\f=u\rfloor d\f+ d(u\rfloor\f), \qquad
\bL_u(\f\w\si)=\bL_u(\f)\w\si +(-1)^{[u][\f]}\f\w\bL_u(\si).
\ee

For instance, let us denote $d_\la\f=\bL_{d_\la}\f$ and
$d_\La=d_{\la_1}\cdots d_{\la_k}$. Given graded functions $f^\La$,
$f'$ and a graded form $\Phi$, there are useful relations
\mar{0606a-c}\ben
&& \op\sum_{0\leq |\La|\leq k} f^\La d_\La f'd^nx= \op\sum_{0\leq
|\La|\leq k} (-1)^{|\La|}d_\La (f^\La) f'd^nx + d_H\si,
\label{0606a} \\
&& \op\sum_{0\leq |\La|\leq k} (-1)^{|\La|}d_\La(f^\La \Phi)=
\op\sum_{0\leq |\La|\leq k} \eta (f)^\La d_\La \Phi, \label{0606b}
\\ && \eta (f)^\La = \op\sum_{0\leq|\Si|\leq
k-|\La|}(-1)^{|\Si+\La|} C^{|\Si|}_{|\Si+\La|} d_\Si f^{\Si+\La},
\qquad C^a_b=\frac{b!}{a!(b-a)!}, \label{0606c}\\
&& (\eta\circ\eta)(f)^\La=f^\La. \nonumber
\een

The BGDA $\cP^*[Q;Y]$ is decomposed into $\cP^0[Q;Y]$-modules
$\cP^{k,r}[Q;Y]$ of $k$-contact and $r$-horizontal graded forms
\be
&& \f=\op\sum_{0\leq|\La_i|}\f^{\La_1\ldots \La_k}_{A_1\ldots A_k
\m_1\ldots\m_r} \th^{A_1}_{\La_1}\w\cdots\w\th^{A_k}_{\La_k}\w
dx^{\m_1}\w\cdots\w
dx^{\m_r}, \qquad \th^A_\La=ds^A_\La -s^A_{\la+\La}dx^\la, \\
&& h_k: \cP^*[Q;Y] \to \cP^{k,*}[Q;Y], \qquad h^r: \cP^*[Q;Y] \to
\cP^{*,r}[Q;Y].
\ee
Accordingly, the graded exterior differential on $\cP^*[Q;Y]$
falls into the sum $d=d_V+d_H$ of the vertical and total
differentials where $d_H\f= dx^\la\w d_\la\f$.

A graded derivation $u$ (\ref{w5}) is called contact if the Lie
derivative $\bL_u$ preserves the ideal of contact graded forms of
the BGDA $\cP^*[Q;Y]$. Further, we restrict our consideration to
vertical contact graded derivations, vanishing on $\cO^*X$. Such a
derivation
\mar{0672}\beq
\vt=\up^A\dr_A + \op\sum_{0<|\La|} d_\La\up^A\dr_A^\La
\label{0672}
\eeq
is determined by its first summand $\up=\up^A\dr_A$, called a
generalized vector field. The relations
\be
\vt\rfloor d_H\f=-d_H(\vt\rfloor\f), \qquad
\bL_\vt(d_H\f)=d_H(\bL_\vt\f), \qquad \f\in\cP^*[Q;Y],
\ee
hold. A vertical contact graded derivation $\vt$ (\ref{0672}) is
called nilpotent if $\bL_\vt(\bL_\vt\f)=0$ for any horizontal
graded form $\f\in \cP^{0,*}[Q;Y]$. One can show that $\vt$
(\ref{0672}) is nilpotent only if it is odd and iff all $\up^A$
obey the equality
\mar{0688}\beq
\vt(\up)=\vt(\up^A\dr_A)=\op\sum_{0\leq|\Si|}
\up^B_\Si\dr^\Si_B(\up^A)\dr_A=0. \label{0688}
\eeq

Lagrangian systems are described in terms of the BGDA $\cP^*[Q;Y]$
and its graded derivations as follows \cite{cmp05,jmp05,jmp05a}.
The differentials $d_H$ and $d_V$, the projector
\be
\vr=\op\sum_{k>0} \frac1k\ol\vr\circ h_k\circ h^n, \qquad
\ol\vr(\f)= \op\sum_{0\leq|\La|} (-1)^{\nm\La}\th^A\w
[d_\La(\dr^\La_A\rfloor\f)], \qquad \f\in \cP^{>0,n}[Q;Y],
\ee
and the graded variational operator $\dl=\vr\circ d$ split the
BGDA $\cP^*[Q;Y]$ into the graded variational bicomplex
\cite{cmp05,jmp05,barn}. We restrict our consideration to its
short variational subcomplex
\mar{g111}\beq
0\to \Bbb R\ar \cP^0[Q;Y]\ar^{d_H}\cP^{0,1}[Q;Y] \cdots \ar^{d_H}
\cP^{0,n}[Q;Y]\ar^\dl \bE_1 =\vr(\cP^{1,n}[Q;Y]). \label{g111}
\eeq
One can think of its even elements
\mar{0709,'}\ben
&& L=\cL d^nx\in \cP^{0,n}[Q;Y],
\label{0709}\\
&& \dl L= \th^A\w \cE_A d^nx=\op\sum_{0\leq|\La|}
 (-1)^{|\La|}\th^A\w d_\La (\dr^\La_A L)d^nx\in \bE_1 \label{0709'}
\een
as being a graded Lagrangian and its Euler--Lagrange operator,
respectively. They possess the following properties.

(i) The complex (\ref{g111}) is exact at all the terms, except
$\Bbb R$. In particular, any $\dl$-closed (i.e., variationally
trivial) graded density $L\in \cP^{0,n}[Q;Y]$ is $d_H$-exact.

(ii) The identity $(\dl\circ\dl)(L)=0$ leads to the useful
equalities
\mar{w51}\beq
 \eta(\dr_B\cE_A)^\La=(-1)^{[A][B]}\dr_A^\La\cE_B. \label{w51}
\eeq

(iii) By virtue of the formula (\ref{0606a}), the Lie derivative
$\bL_\vt L$ of a Lagrangian $L$ along a vertical contact graded
derivation $\vt$ (\ref{0672}) admits the decomposition
\mar{g107}\beq
\bL_\vt L= \up\rfloor\dl L +d_H\si. \label{g107}
\eeq
One says that an odd vertical contact graded derivation $\vt$
(\ref{0672}) is a variational supersymmetry of a Lagrangian $L$ if
the Lie derivative $\bL_\vt L$ is $d_H$-exact or, equivalently,
the odd graded density $\up\rfloor\dl L=\up^A\cE_Ad^nx$ is
$d_H$-exact.

For the sake of simplicity, the common symbol $\up$ further stands
for a generalized vector field $\up$, the vertical contact graded
derivation $\vt$ (\ref{0672}) determined by $\up$  and the Lie
derivative $\bL_\vt$. We agree to call all these operators a
graded derivation of the BGDA $\cP^*[Q;Y]$.

One also deals with right contact graded derivations $\op\up^\lto
={\op\dr^\lto}_A\up^A$ of the BGDA $\cP^*[Q;Y]$. They act on
graded forms $\f$ on the right by the rule
\be
\op\up^\lto(\f)=\op d^\lto(\f)\lfloor \op\up^\lto +\op
d^\lto(\f\lfloor\op\up^\lto), \qquad
\op\up^\lto(\f\w\f')=(-1)^{[\f'][\op\up^\lto]}\op\up^\lto(\f)\w\f'+
\f\w\op\up^\lto(\f').
\ee
For instance,
\be
{\op\dr^\lto}_A(\f)=(-1)^{([\f]+1)[A]}\dr_A(\f), \qquad {\op
d^\lto}_\La=d_\La, \qquad {\op d^\lto}_H(\f)= (-1)^{|\f|}d_H(\f).
\ee
With right graded derivations, one obtains the right
Euler--Lagrange operator
\be
\op\dl^\lto L= {\op\cE^\lto}_Ad^nx\w \th^A, \qquad {\op\cE^\lto}_A
=\op\sum_{0\leq|\La|}
 (-1)^{|\La|}d_\La (\rdr^\La_A (L)).
\ee
An odd right graded derivation $\op\up^\lto$ is a variational
supersymmetry of a Lagrangian $L$ iff the odd graded density
${\op\cE^\lto}_A\up^Ad^nx$ is $d_H$-exact.

\bigskip
\bigskip

\noindent {\bf III. NOETHER IDENTITIES AND ANTIFIELDS}
\bigskip

Given a graded Lagrangian $L$ (\ref{0709}), one can associate to
its Euler--Lagrange operator $\dl L$ (\ref{0709'}) the exact chain
Koszul--Tate complex with the boundary operator whose nilpotency
conditions provide the Noether and higher-stage Noether identities
for $\dl L$ \cite{jmp05a}.

Let us start with the following notation. Given vector bundles $V,
V', E, E'$ over $X$, we consider the BGDA
\be
\cP^*[V'V;Q;Y;EE']= \cP^*[V'\op\times_X V\op\times_XQ;Y\op\times_X
E\op\times_X E'].
\ee
By a density-dual of a vector bundle $E\to X$ is meant
\be
\ol E^*=E^*\ot_X\op\w^n T^*X.
\ee
By $\ol Y^*$ is denoted the density-dual of the vector bundle
which an affine bundle $Y$ is modelled on.

Let us extend the BGDA $\cP^*[Q;Y]$ to the BGDA $\cP^*[\ol
Y^*;Q;Y;\ol Q^*]$ whose basis is $\{s^A, \ol s_A\}$, where $[\ol
s_A]=([A]+1){\rm mod}\,2$. We call $\ol s_A$ the antifields of
antifield number Ant$[\ol s_A]= 1$. The BGDA $\cP^*[\ol
Y^*;Q;Y;\ol Q^*]$ is provided with the nilpotent right graded
derivation $\ol\dl=\rdr^A \cE_A$, where $\cE_A$ are the graded
variational derivatives (\ref{0709'}). We call $\ol\dl$ the
Koszul--Tate differential, and say that an element $\f\in
\cP^*[\ol Y^*;Q;Y;\ol Q^*]$ vanishes on the shell if it is
$\ol\dl$-exact, i.e., $\f=\ol\dl\si$. Let us consider the module
$\cP^{0,n}[\ol Y^*;Q;Y;\ol Q^*]$ of graded densities. It contains
the chain complex
\mar{v042}\beq
0\lto \im\ol\dl \llr^{\ol\dl} \cP^{0,n}[\ol Y^*;Q;Y;\ol Q^*]_1
\llr^{\ol\dl} \cP^{0,n}[\ol Y^*;Q;Y;\ol Q^*]_2. \label{v042}
\eeq
This complex is exact at $\im\ol\dl$. Let us consider its first
homology $H_1(\ol \dl)$.

A generic one-chain of the complex (\ref{v042}) takes the form
\mar{0712}\beq
\Phi= \op\sum_{0\leq|\La|} \Phi^{A,\La}\ol s_{\La A} d^nx, \qquad
\Phi^{A,\La}\in \cP^0[Q;Y]. \label{0712}
\eeq
Then the cycle condition $\ol\dl \Phi=0$ provides a reduction
condition
\mar{0713}\beq
\op\sum_{0\leq|\La|} \Phi^{A,\La} d_\La \cE_A d^nx=0 \label{0713}
\eeq
on the graded variational derivatives $\cE_A$. Conversely, any
reduction condition of form (\ref{0713}) comes from some cycle
(\ref{0712}). The reduction condition (\ref{0713}) is trivial  if
a cycle is a boundary, i.e., it takes the form
\be
\Phi= \op\sum_{0\leq|\La|,|\Si|} T^{(A\La)(B\Si)}d_\Si\cE_B\ol
s_{\La A}d^nx, \qquad T^{(A\La)(B\Si)}=-(-1)^{[A][B]}
T^{(B\Si)(A\La)}.
\ee
A Lagrangian system is called degenerate if there exist nontrivial
reduction conditions (\ref{0713}), called the Noether identities.

One can say something more if the $\cP^0[Q;Y]$-module $H_1(\ol
\dl)$ is finitely generated, i.e., there are elements of $H_1(\ol
\dl)$ making up a free graded $C^\infty(X)$-module $\cC_{(0)}$ of
finite rank. By virtue of the Serre--Swan theorem \cite{jmp05a},
$\cC_{(0)}$ is isomorphic to the module of sections of the product
$\ol V^*\op\times_X \ol E^*$ where $\ol V^*$, $\ol E^*$ are the
density-duals of some vector bundles $V\to X$ and $E\to X$. Let
$\{\Delta_r\}$ be a basis for this $C^\infty(X)$-module. Every
element $\Phi\in H_1(\ol \dl)$ factorizes
\mar{v63}\beq
\Phi= \op\sum_{0\leq|\Xi|} G^{r,\Xi} d_\Xi \Delta_rd^nx, \qquad
\Delta_r=\op\sum_{0\leq|\La|} \Delta_r^{A,\La}\ol s_{\La A},
\label{v63}
\eeq
via elements  $\Delta_r\in \cC_{(0)}$, called the Noether
operators. Accordingly, any Noether identity (\ref{0713}) is a
corollary of the identities
\mar{v64}\beq
 \op\sum_{0\leq|\La|} \Delta_r^{A,\La} d_\La \cE_A=0,
\label{v64}
\eeq
called the complete Noether identities. In this case, the complex
(\ref{v042}) can be extended to a one-exact complex with a
boundary operator whose nilpotency conditions are equivalent to
the Noether identities (\ref{v64}).

Note that, if there is no danger of confusion, elements of
homology are identified to its representatives. A chain complex is
called $r$-exact if its homology of degree $k\leq r$ is trivial.

Let us enlarge the BGDA $\cP^*[\ol Y^*;Q;Y;\ol Q^*]$ to the BGDA
$\cP^*[\ol E^*\ol Y^*;Q;Y;\ol Q^*\ol V^*]$ possessing the basis
$\{s^A,\ol s_A, \ol c_r\}$ where $[\ol c_r]=([\Delta_r]+1){\rm
mod}\,2$ and Ant$[\ol c_r]=2$. This BGDA is provided with the
nilpotent right graded derivation $\dl_0=\ol\dl + \rdr^r\Delta_r$.
Its nilpotency conditions (\ref{0688}) are equivalent to the
Noether identities (\ref{v64}). Then the module $\cP^{0,n}[\ol
E^*\ol Y^*;Q;Y;\ol Q^*\ol V^*]_{\leq 3}$ of graded densities of
antifield number Ant$[\f]\leq 3$ is split into the chain complex
\mar{v66}\ben
&&0\lto \im\ol\dl \llr^{\ol\dl} \cP^{0,n}[\ol Y^*;Q;Y;\ol
Q^*]_1\llr^{\dl_0}
\cP^{0,n}[\ol E^*\ol Y^*;Q;Y;\ol Q^*\ol V^*]_2 \label{v66}\\
&& \qquad \llr^{\dl_0} \cP^{0,n}[\ol E^*\ol Y^*;Q;Y;\ol Q^*\ol
V^*]_3. \nonumber
\een
Let $H_*(\dl_0)$ be its homology. We have
$H_0(\dl_0)=H_0(\ol\dl)=0$. Furthermore, any one-cycle $\Phi$ up
to a boundary takes the form (\ref{v63}) and, therefore, it is a
$\dl_0$-boundary
\be
\Phi= \op\sum_{0\leq|\Si|} G^{r,\Xi} d_\Xi \Delta_rd^nx
=\dl_0(\op\sum_{0\leq|\Si|} G^{r,\Xi}\ol c_{\Xi r}d^nx).
\ee
Hence, the complex (\ref{v66}) is one-exact.

Turn now to the homology $H_2(\dl_0)$ of the complex (\ref{v66}).
A generic two-chain  reads
\mar{v77}\beq
\Phi= G + H= \op\sum_{0\leq|\La|} G^{r,\La}\ol c_{\La r}d^nx +
\op\sum_{0\leq|\La|,|\Si|} H^{(A,\La)(B,\Si)}\ol s_{\La A}\ol
s_{\Si B}d^nx. \label{v77}
\eeq
The cycle condition $\dl_0 \Phi=0$ provides the reduction
condition
\mar{v79}\beq
 \op\sum_{0\leq|\La|} G^{r,\La}d_\La\Delta_rd^nx +\ol\dl H=0
\label{v79}
\eeq
on the Noether operators (\ref{v63}). Conversely, let
\be
\Phi=\op\sum_{0\leq|\La|} G^{r,\La}\ol c_{\La r}d^nx\in
\cP^{0,n}[\ol E^*\ol Y^*;Q;Y;\ol Q^*\ol V^*]_2
\ee
be a graded density such that the reduction condition (\ref{v79})
holds. Obviously, this reduction condition is a cycle condition of
the two-chain (\ref{v77}). It is trivial either if a two-cycle
$\Phi$ (\ref{v77}) is a boundary or its summand $G$, linear in
antifields, vanishes on the shell.

A degenerate Lagrangian system is said to be one-stage reducible
if there exist nontrivial reduction conditions (\ref{v79}), called
the first-stage Noether identities. One can show that first-stage
Noether identities are identified to elements of the homology
$H_2(\dl_0)$ iff any $\ol\dl$-cycle $\f\in \cP^{0,n}[\ol
Y^*;Q;Y;\ol Q^*]_2$ is a $\dl_0$-boundary \cite{jmp05a}.
Furthermore, if first-stage Noether identities are finitely
generated, the complex (\ref{v66}) is extended to a two-exact
complex with a boundary operator whose nilpotency conditions are
equivalent to complete Noether and first-stage Noether identities.
If the third homology of this complex is not trivial, we have
second-stage Noether identities, and so on. Iterating the
arguments, one says that a degenerate Lagrangian system
$(\cP^*[Q;Y],L)$ is $N$-stage reducible if the following holds
\cite{jmp05a}.

(i) There exist vector bundles $V_1,\ldots, V_N, E_1, \ldots, E_N$
over $X$, and the BGDA $\cP^*[Q;Y]$ is enlarged to the BGDA
\mar{v91}\beq
\ol\cP^*\{N\}=\cP^*[\ol E^*_N\cdots\ol E^*_1\ol E^*\ol Y^*;Q;Y;\ol
Q^*\ol V^*\ol V^*_1\cdots\ol V_N^*] \label{v91}
\eeq
with the basis $\{s^A,\ol s_A, \ol c_r, \ol c_{r_1}, \ldots, \ol
c_{r_N}\}$ graded by antifield numbers Ant$[\ol c_{r_k}]=k+2$. Let
the indexes $k=-1,0$ further stand for $\ol s_A$ and $\ol c_r$,
respectively.

(ii) The BGDA $\ol\cP^*\{N\}$ (\ref{v91}) is provided with a
nilpotent graded derivation
\mar{v92,'}\ben
&& \dl_N=\rdr^A \cE_A + \op\sum_{0\leq|\La|} \rdr^r
\Delta_r^{A,\La}\ol s_{\La A} + \op\sum_{1\leq k\leq N}\rdr^{r_k}
\Delta_{r_k}, \label{v92}\\
&& \Delta_{r_k}=G_{r_k} + h_{r_k}= \op\sum_{0\leq|\La|}
\Delta_{r_k}^{r_{k-1},\La}\ol c_{\La r_{k-1}} + \op\sum_{0\leq
|\Si|, |\Xi|}(h_{r_k}^{(r_{k-2},\Si)(A,\Xi)}\ol c_{\Si r_{k-2}}\ol
s_{\Xi A}+...), \label{v92'}
\een
of antifield number -1. It is called the $N$-stage Koszul--Tate
differential.

(iii) With $\dl_N$, the module $\ol\cP^{0,n}\{N\}_{\leq N+3}$ of
graded densities of antifield number Ant$[\f]\leq N+3$ is split
into the $(N+2)$-exact chain complex
\mar{v94}\ben
&&0\lto \im \ol\dl \llr^{\ol\dl} \cP^{0,n}[\ol Y^*;Q;Y;\ol
Q^*]_1\llr^{\dl_0} \ol\cP^{0,n}\{0\}_2\llr^{\dl_1}
\ol\cP^{0,n}\{1\}_3\cdots
\label{v94}\\
&& \qquad
 \llr^{\dl_{N-1}} \ol\cP^{0,n}\{N-1\}_{N+1}
\llr^{\dl_N} \ol\cP^{0,n}\{N\}_{N+2}\llr^{\dl_N}
\ol\cP^{0,n}\{N\}_{N+3}, \nonumber
\een
which satisfies the homology regularity condition. This condition
states that, any $\dl_{k<N-1}$-cycle $\f\in
\ol\cP^{0,n}\{k\}_{k+3}\subset \ol\cP^{0,n}\{k+1\}_{k+3}$ is a
$\dl_{k+1}$-boundary.

(iv) The nilpotency of $\dl_N$ implies the Noether identities
(\ref{v64}) and the $k$-stage Noether identities
\mar{v93}\beq
\op\sum_{0\leq|\La|} \Delta_{r_k}^{r_{k-1},\La}d_\La
(\op\sum_{0\leq|\Si|} \Delta_{r_{k-1}}^{r_{k-2},\Si}\ol c_{\Si
r_{k-2}}) + \ol\dl(\op\sum_{0\leq |\Si|,
|\Xi|}h_{r_k}^{(r_{k-2},\Si)(A,\Xi)}\ol c_{\Si r_{k-2}}\ol s_{\Xi
A})=0  \label{v93}
\eeq
for $k=1,\ldots,N$. Accordingly, $\Delta_{r_k}$ (\ref{v92'}) are
called the $k$-stage Noether operators.

\bigskip
\bigskip

\noindent {\bf IV. GAUGE SUPERSYMMETRIES AND GHOSTS}
\bigskip

The Noether second theorem associates to the Koszul--Tate complex
(\ref{v94}) the sequence (\ref{w36}) graded in ghosts. Its ascent
operator $\up_e$ (\ref{w108}) provides the gauge and higher-stage
gauge supersymmetries of an original Lagrangian $L$.

Given the BGDA $\ol\cP^*\{N\}$ (\ref{v91}), let us consider the
BGDA
\mar{w5'}\beq
\cP^*\{N\}=\cP^*[V_N\cdots V_1V;Q;Y;EE_1\cdots E_N] \label{w5'}
\eeq
with the basis $\{s^A, c^r, c^{r_1}, \ldots, c^{r_N}\}$ and the
BGDA
\mar{w6'}\beq
P^*\{N\}=\cP^*[\ol E^*_N\cdots\ol E^*_1\ol E^* V_N\cdots V_1V\ol
Y^*;Q;Y;\ol Q^*EE_1\cdots E_N\ol V^*\ol V^*_1\cdots\ol V_N^*]
\label{w6'}
\eeq
with the basis
\mar{w7}\beq
\{s^A, c^r, c^{r_1}, \ldots, c^{r_N},\ol s_A,\ol c_r, \ol c_{r_1},
\ldots, \ol c_{r_N}\},\label{w7}
\eeq
where $[c^{r_k}]=([\ol c_{r_k}]+1){\rm mod}\,2$ and
Ant$[c^{r_k}]=-(k+1)$. We call $c^{r_k}$, $k=0,\ldots,N$, the
ghosts of ghost number gh$[c^{r_k}]=k+1$. Clearly, the BGDAs
$\ol\cP^*\{N\}$ (\ref{v91}) and $\cP^*\{N\}$ (\ref{w5'}) are
subalgebras of the BGDA $P^*\{N\}$ (\ref{w6'}). The $N$-stage
Koszul--Tate differential $\dl_N$ (\ref{v92}) is naturally
prolonged to a graded derivation of the BGDA $P^*\{N\}$
(\ref{w6'}).

Let us extend an original Lagrangian $L$ to the even graded
density
\mar{w8}\beq
L_e=\cL_ed^nx=L+L_1=L + \op\sum_{0\leq k\leq N}
c^{r_k}\Delta_{r_k}d^nx=L +\dl_N( \op\sum_{0\leq k\leq N}
c^{r_k}\ol c_{r_k}d^nx), \label{w8}
\eeq
of zero antifield number where $L_1$ is linear in ghosts. It is
readily observed that $\dl_N(L_e)=0$, i.e., $\dl_N$ is a
variational supersymmetry of the Lagrangian $L_e$ (\ref{w8}). It
follows that
\mar{w16}\ben
 && [\frac{\op\dl^\lto \cL_e}{\dl \ol
s_A}\cE_A +\op\sum_{0\leq k\leq N} \frac{\op\dl^\lto \cL_e}{\dl
\ol c_{r_k}}\Delta_{r_k}]d^nx = [\up^A\cE_A + \op\sum_{0\leq k\leq
N}\up^{r_k}\frac{\dl
\cL_e}{\dl c^{r_k}}]d^nx= d_H\si, \label{w16}\\
&& \up^A= \frac{\op\dl^\lto \cL_e}{\dl \ol s_A}=u^A+w^A
=\op\sum_{0\leq|\La|} c^r_\La\eta(\Delta^A_r)^\La + \op\sum_{1\leq
i\leq N}\op\sum_{0\leq|\La|}
c^{r_i}_\La\eta(\op\dr^\lto{}^A(h_{r_i}))^\La, \nonumber\\
&& \up^{r_k}=\frac{\op\dl^\lto \cL_e}{\dl \ol c_{r_k}} =u^{r_k}+
w^{r_k}= \op\sum_{0\leq|\La|}
c^{r_{k+1}}_\La\eta(\Delta^{r_k}_{r_{k+1}})^\La
+\op\sum_{k+1<i\leq N} \op\sum_{0\leq|\La|}
c^{r_i}_\La\eta(\op\dr^\lto{}^{r_k}(h_{r_i}))^\La, \nonumber
\een
(see the formulas (\ref{0606b}) -- (\ref{0606c})). The equality
(\ref{w16}) falls into the set of equalities
\mar{w19,20}\ben
&& \frac{\op\dl^\lto (c^r\Delta_r)}{\dl \ol s_A}\cE_Ad^nx
=u^A\cE_Ad^nx=d_H\si_0, \label{w19}\\
&&  [\frac{\op\dl^\lto (c^{r_i}\Delta_{r_i})}{\dl \ol s_A}\cE_A
+\op\sum_{k<i} \frac{\op\dl^\lto (c^{r_i}\Delta_{r_i})}{\dl \ol
c_{r_k}}\Delta_{r_k}]d^nx= d_H\si_i, \qquad i=1,\ldots,N.
\label{w20}
\een
It follows from the equality (\ref{w19}) and formula (\ref{g107})
that the graded derivation
\mar{w33}\beq
u= u^A\frac{\dr}{\dr s^A}, \qquad u^A =\op\sum_{0\leq|\La|}
c^r_\La\eta(\Delta^A_r)^\La, \label{w33}
\eeq
is a variational supersymmetry of an original Lagrangian $L$. This
variational supersymmetry is parameterized by ghosts $c^r$, i.e.,
it is a gauge supersymmetry of $L$ \cite{cmp05,jmp05}. Every
equality (\ref{w20}) is split into a set of equalities with
respect to the polynomial degree in antifields. Let us consider
the one, linear in antifields $\ol c_{r_{i-2}}$ and their jets
(where by $\ol c_{r_{-1}}$ are meant $\ol s_A$). It is brought
into the form
\be
[\op\sum_{0\leq|\Xi|}
(-1)^{|\Xi|}d_\Xi(c^{r_i}\op\sum_{0\leq|\Si|}
h_{r_i}^{(r_{i-2},\Si)(A,\Xi)} \ol c_{\Si r_{i-2}})\cE_A +
u^{r_{i-1}}\op\sum_{0\leq|\Xi|} \Delta_{r_{i-1}}^{r_{i-2},\Xi}\ol
c_{\Xi r_{i-2}}]d^nx= d_H\si_i.
\ee
Using the relation (\ref{0606a}), we obtain
\be
[\op\sum_{0\leq|\Xi|} c^{r_i}\op\sum_{0\leq|\Si|}
h_{r_i}^{(r_{i-2},\Si)(A,\Xi)} \ol c_{\Si r_{i-2}} d_\Xi\cE_A +
u^{r_{i-1}}\op\sum_{0\leq|\Xi|} \Delta_{r_{i-1}}^{r_{i-2},\Xi}\ol
c_{\Xi r_{i-2}}]d^nx= d_H\si'_i.
\ee
The variational derivative of the both sides of this equality with
respect to the antifield $\ol c_{r_{i-2}}$ leads to the relation
\be
\op\sum_{0\leq|\Si|} \eta(h_{r_i}^{(r_{i-2})(A,\Xi)})^\Si
d_\Si(c^{r_i} d_\Xi\cE_A) + \op\sum_{0\leq|\Si|}
u^{r_{i-1}}_\Si\eta (\Delta^{r_{i-2}}_{r_{i-1}})^\Si=0,
\ee
which takes the form
\mar{w34}\beq
\op\sum_{0\leq|\Si|} d_\Si u^{r_{i-1}}\frac{\dr}{\dr
c^{r_{i-1}}_\Si} u^{r_{i-2}} =\ol\dl(\al^{r_{i-2}}), \qquad
\al^{r_{i-2}} = -\op\sum_{0\leq|\Si|}
\eta(h_{r_i}^{(r_{i-2})(A,\Xi)})^\Si d_\Si(c^{r_i} \ol s_{\Xi A}).
\label{w34}
\eeq
Therefore, the odd graded derivations
\mar{w38}\beq
u_{(k)}= u^{r_{k-1}}\frac{\dr}{\dr c^{r_{k-1}}}, \qquad
u^{r_{k-1}}=\op\sum_{0\leq|\La|}
c^{r_k}_\La\eta(\Delta^{r_{k-1}}_{r_k})^\La, \qquad k=1,\ldots,N,
\label{w38}
\eeq
are the $k$-stage gauge supersymmetries \cite{jmp05}. The graded
derivations $u$ (\ref{w33}), $u_{(k)}$ (\ref{w38}) are assembled
into the graded derivation
\mar{w108}\beq
u_e=u + \op\sum_{1\leq k\leq N} u_{(k)} \label{w108}
\eeq
of ghost number 1, that we agree to call the total gauge operator.
It is an ascent operator of the sequence
\mar{w36}\beq
0\to \cS^0[Q;Y]\ar^{u_e} \cP^0\{N\}_1\ar^{u_e}
\cP^0\{N\}_2\ar^{u_e}\cdots. \label{w36}
\eeq

The total gauge operator (\ref{w108}) need not be nilpotent. One
can say that gauge and higher-stage gauge supersymmetries of a
Lagrangian system form an algebra on the shell if the graded
derivation (\ref{w108}) can be extended to a graded derivation
$u_E$ of ghost number 1 by means of terms of higher polynomial
degree in ghosts, and $u_E$ is nilpotent on the shell. Namely, we
have
\mar{w109}\beq
u_E=u_e+ \xi= u^A\dr_A + \op\sum_{1\leq k\leq N}(u^{r_{k-1}}
+\xi^{r_{k-1}})\dr_{r_{k-1}} +\xi^{r_N}\dr_{r_N}, \label{w109}
\eeq
where all the coefficients $\xi^{r_k}$, $k=0,\ldots,N$, are at
least quadratic in ghosts and $(u_E\circ u_E)(f)$ is
$\ol\dl$-exact for any graded function $f\in \cP^0\{N\}\subset
P^0\{N\}$. This nilpotency condition falls into a set of
equalities with respect to the polynomial degree in ghosts. Let us
write the first and second of them involving the coefficients
$\xi_2^{r_{k-1}}$ quadratic in ghosts. We have
\mar{w110,3}\ben
&& \op\sum_{0\leq|\Si|} d_\Si u^r\dr^\Si_r u^B=\ol\dl(\al^B_1),
\qquad \op\sum_{0\leq|\Si|} d_\Si u^{r_{k-1}}\dr_{r_{k-1}}^\Si
u^{r_{k-2}}
=\ol\dl(\al^{r_{k-2}}_1), \quad 2\leq k\leq N, \label{w110}\\
&&  \op\sum_{0\leq|\Si|}[d_\Si u^A\dr^\Si_A u^B
+d_\Si\xi^r_2\dr^\Si_r u^B]=\ol\dl(\al^B_2),
\label{w111} \\
&& \op\sum_{0\leq|\Si|}[d_\Si u^A\dr^\Si_A u^{r_{k-1}} +
d_\Si\xi^{r_k}_2\dr^\Si_{r_k} u^{r_{k-1}} + d_\Si
u^{r'_{k-1}}\dr^\Si_{r'_{k-1}}\xi^{r_{k-1}}_2]=
\ol\dl(\al^{r_{k-1}}_2),
 \label{w112}\\
&& \xi_2^r=\xi^{r,\La,\Si}_{r',r''} c^{r'}_\La c^{r''}_\Si, \qquad
\xi_2^{r_k}=\xi^{r_k,\La,\Si}_{r,r'_k} c^r_\La c^{r'_k}_\Si,
\qquad 2\leq k\leq N. \label{w113}
\een
The equalities (\ref{w110}) reproduce the relations (\ref{w34}).
The equalities (\ref{w111}) -- (\ref{w112}) provide the
generalized commutation relations on the shell between gauge and
higher-stage gauge supersymmetries, and one  can think of the
coefficients $\xi_2$ (\ref{w113}) as being {\it sui generis}
generalized structure functions \cite{jmp05,fulp02}.

\bigskip
\bigskip

\noindent {\bf V. THE MASTER EQUATION AND EXTENDED LAGRANGIAN}
\bigskip

The BGDA $P^*\{N\}$ (\ref{w6'}) with the basis (\ref{w7})
exemplifies Lagrangian systems of the following particular type.

Let $Y_0\to X$ be an affine bundle and $\ol Y^*_0$ the
density-dual of the vector bundle which an affine bundle $Y_0$ is
modelled on. Let $Y_1\to X$ be a vector bundle and $\ol Y_1^*$ its
density-dual. We consider the BGDA $\cP^*[\ol Y^*_0;Y_1;Y_0;\ol
Y_1^*]$ endowed with the basis $\{y^a,\ol y_a\}$, $[\ol
y_a]=([y^a]+1){\rm mod}\,2$, whose elements $y^a$ and $\ol y_a$
are called the fields and antifields, respectively. Then one can
associate to any Lagrangian
\mar{w40}\beq
\gL d^nx\in \cP^{0,n}[\ol Y^*_0;Y_1;Y_0;\ol Y_1^*] \label{w40}
\eeq
the odd graded derivations
\mar{w37}\beq
\up=\op\cE^\lto{}^a\dr_a=\frac{\op\dl^\lto \gL}{\dl \ol y_a}
\frac{\dr}{\dr y^a}, \qquad \ol\up=
\rdr^a\cE_a=\frac{\op\dr^\lto}{\dr \ol y_a}\frac{\dl \gL}{\dl y^a}
\label{w37}
\eeq
of the BGDA $\cP^*[\ol Y^*_0;Y_1;Y_0;\ol Y_1^*]$.

\begin{theo} \label{w39} \mar{w39} The following conditions are
equivalent:

(i) the graded derivation $\up$ (\ref{w37}) is a variational
supersymmetry of a Lagrangian $\gL d^nx$ (\ref{w40}),

(ii) the graded derivation $\ol\up$ (\ref{w37}) is a variational
supersymmetry of $\gL d^nx$ (\ref{w40}),

(iii) the composition $(\up-\ol\up)\circ(\up +\ol\up)$ acting on
even graded functions $f\in \cP^0[\ol Y^*_0;Y_1;Y_0;\ol Y_1^*]$
(or, equivalently, $(\up+\ol\up)\circ(\up -\ol\up)$ acting on the
odd ones) vanishes.
\end{theo}

\begin{proof} By virtue of the formula
(\ref{g107}), the conditions (i) and (ii) are equivalent to the
equality
\mar{w44}\beq
\op\cE^\lto{}^a\cE_ad^nx=\frac{\op\dl^\lto \gL}{\dl \ol
y_a}\frac{\dl \gL}{\dl y^a}d^nx =d_H\si, \label{w44}
\eeq
called the master equation. The equality (\ref{w44}), in turn, is
equivalent to the condition that the odd graded density
$\op\cE^\lto{}^a\cE_a d^nx$ is variationally trivial. Replacing
the right variational derivatives $\op\cE^\lto{}^a$ in the
equality (\ref{w44}) with the left ones $(-1)^{[a]+1}\cE^a$. We
obtain
\be
\op\sum_a (-1)^{[a]}\cE^a\cE_a d^nx=d_H\si.
\ee
The variational operator acting on this relation leads to the
equalities
\be
&&
\op\sum_{0\leq|\La|}(-1)^{[a]+|\La|}d_\La(\dr^\La_b(\cE^a\cE_a))=
\op\sum_{0\leq|\La|}(-1)^{[a]}[\eta(\dr_b\cE^a)^\La\cE_{\La a} +
\eta(\dr_b\cE_a)^\La\cE^a_\La)]=0, \\
&& \op\sum_{0\leq|\La|}(-1)^{[a]+|\La|}d_\La(\dr^{\La
b}(\cE^a\cE_a)) =
\op\sum_{0\leq|\La|}(-1)^{[a]}[\eta(\dr^b\cE^a)^\La\cE_{\La a} +
\eta(\dr^b\cE_a)\cE^a_\La] = 0.
\ee
Due to the formulas (\ref{w51}), these equalities are brought into
the form
\mar{z1,2}\ben
&& \op\sum_{0\leq|\La|}(-1)^{[a]}[(-1)^{[b]([a]+1)}\dr^{\La
a}\cE_b\cE_{\La
a} + (-1)^{[b][a]}\dr_a^\La\cE_b\cE^a_\La]=0, \label{z1}\\
&& \op\sum_{0\leq|\La|}(-1)^{[a]}[(-1)^{([b]+1)([a]+1)}\dr^{\La
a}\cE^b\cE_{\La a} + (-1)^{([b]+1)[a]}\dr_a^\La\cE^b\cE^a_\La]=0
\label{z2}
\een
for all $\cE_b$ and $\cE^b$. Returning to the right variational
derivatives, we obtain the relations
\mar{z20}\beq
\op\dr^\lto{}^{\La a}(\cE_b)\cE_{\La a} +
(-1)^{[b]}\op\cE^\lto{}^a_\La \dr^\La_a\cE_b=0, \qquad
\op\cE^\lto{}^a_\La \dr^\La_a\op\cE^\lto{}^b + (-1)^{[b]+1}
\op\dr^\lto{}^{\La a}(\op\cE^\lto{}^b)\cE_{\La a}=0. \label{z20}
\eeq
A direct computation shows that they are equivalent to the
condition (iii).
\end{proof}

It is readily observed that the equalities (\ref{z1}) --
(\ref{z2}) and, equivalently, the equalities (\ref{z20}) are
Noether identities of a Lagrangian (\ref{w40}).

Note that any variationally trivial Lagrangian satisfies the
master equation. We say that a solution of the master equation is
not trivial if both the graded derivations (\ref{w37}) are not
zero. It is readily observed that, if a Lagrangian $\gL d^nx$
provides a nontrivial solution of the master equation and $L_0$ is
a variationally trivial Lagrangian, the sum $\gL d^nx +L_0$ is
also a nontrivial solution of the master equation.

Let us return to an original graded Lagrangian system
$(\cP^*[Q;Y],L)$ and its extension $(P^*\{N\},L_e,u_e)$ to ghosts
and antifields, together with the odd graded derivations
(\ref{w37}) which read
\be
&& \up_e= \vt +\vt_e= \frac{\op\dl^\lto \cL_1}{\dl \ol
s_A}\frac{\dr}{\dr s^A} + \op\sum_{0\leq k\leq N}
\frac{\op\dl^\lto \cL_1}{\dl \ol
c_{r_k}}\frac{\dr}{\dr c^{r_k}},  \\
&& \ol\up_e=\ol\vt +\dl_N  = \frac{\rdr }{\dr \ol
s_A}\frac{\dl\cL_1}{\dl s^A} + [\frac{\rdr }{\dr \ol
s_A}\frac{\dl\cL}{\dl s^A} +\op\sum_{0\leq k\leq N} \frac{\rdr
}{\dr \ol c_{r_k}}\frac{\dl \cL_1}{\dl c^{r_k}}].
\ee
An original Lagrangian provides a trivial solution of the master
equation. A problem is to extend the Lagrangian $L_e$ (\ref{w8})
to a solution of the master equation
\mar{w61}\beq
L_e+L'=L+L_1+L_2+\cdots \label{w61}
\eeq
by means of even terms $L_i$ of zero antifield number and
polynomial degree $i>1$ in ghosts.

\begin{theo} \label{w120} \mar{w120}
The Lagrangian $L_e$ (\ref{w8}) can be extended to a solution
(\ref{w61}) of the master equation only if the total gauge
operator $u_e$ (\ref{w108}) is extended to a graded derivation,
nilpotent on the shell. This extension is independent of
antifields.
\end{theo}

\begin{proof}
Given a Lagrangian (\ref{w61}), the corresponding graded
derivations (\ref{w37}) read
\mar{w102,3}\ben
&& \up=  \frac{\op\dl^\lto (\cL_1+\cL')}{\dl \ol
s_A}\frac{\dr}{\dr s^A} + \op\sum_{0\leq k\leq N}
\frac{\op\dl^\lto (\cL_1+\cL')}{\dl \ol
c_{r_k}}\frac{\dr}{\dr c^{r_k}}, \label{w102} \\
&& \ol\up= \frac{\rdr }{\dr \ol s_A}\frac{\dl(\cL+\cL_1+\cL')}{\dl
s^A} + \op\sum_{0\leq k\leq N} \frac{\rdr }{\dr \ol
c_{r_k}}\frac{\dl (\cL_1+\cL')}{\dl c^{r_k}}. \label{w103}
\een
Then the condition (iii) of Theorem \ref{w39} can be written in
the form (\ref{0688}) as
\mar{w131}\beq
(\up +\ol\up)(\up)=0, \qquad (\up +\ol\up)(\ol\up)=0. \label{w131}
\eeq
It falls into a set of equalities with respect to the polynomial
degree in antifields. Let us put
\be
\up=\up^0+\up^1 +\up', \qquad \ol\up=\ol\up^0 +\ol\up',
\ee
where $\up^0$ and $\up^1$ are the parts of $\up$ (\ref{w102}) of
zero and first polynomial degree in antifields, respectively, and
$\ol\up^0$ is that of $\ol\up$ (\ref{w103}) independent of
antifields. It is readily observed that $\ol\up^0=\ol\dl$ is the
Koszul--Tate differential. Let us consider the part of the first
equality (\ref{w131}) which is independent of antifields. It reads
\mar{w104}\beq
\up^0(\up^0) +\ol\up^0(\up^1)= \up^0(\up^0) +\ol\dl(\up^1)=0,
\label{w104}
\eeq
i.e., the graded derivation $\up^0$ is nilpotent on the shell.
Since the part of $\up^0$ linear in ghosts is exactly the total
gauge operator $u_e$ (\ref{w108}), the graded derivation $\up^0$
provides a desired extension $u_E=\up^0$ (\ref{w109}) of $u_e$
which is nilpotent on the shell.
\end{proof}

Theorem \ref{w120} shows that the Lagrangian  $L_e$ (\ref{w8}) is
extended to a solution of the master equation only if the gauge
and higher-stage gauge supersymmetries of an original Lagrangian
$L$ form an algebra on the shell.

In order to formulate the sufficient condition, let us assume that
the gauge and higher-stage gauge supersymmetries of an original
Lagrangian $L$ form an algebra, i.e., we have a nilpotent
extension of the total gauge operator $u_e$ (\ref{w108}).

\begin{theo} \label{w130} \mar{w130}
Let the total gauge operator $u_e$ (\ref{w108}) admit a nilpotent
extension $u_E$ (\ref{w109}) independent of antifields. Then the
extended Lagrangian
\mar{w133}\beq
L_E=L_e + \op\sum_{0\leq k\leq N}\xi^{r_k}\ol c_{r_k}d^nx
\label{w133}
\eeq
satisfies the master equation.
\end{theo}

\begin{proof}
If the graded derivation $\up^0=u_E$ is nilpotent, then
$\ol\dl(\up^1)=0$ by virtue of the equation (\ref{w104}). It
follows that the part $L^2_1$ of the Lagrangian $L_e$ quadratic in
antifields obeys the relations
$\ol\dl(\op\dl^\lto{}^{r_k}(\cL^2_1))=0$ for all indices $r_k$.
This part consists of the terms $h_{r_k}^{(r_{k-2},\Si)(A,\Xi)}\ol
c_{\Si r_{k-2}}\ol s_{\Xi A}$ (\ref{v92'}), which consequently are
$\ol\dl$-closed. Then the summand $G_{r_k}$ of each cocycle
$\Delta_{r_k}$ (\ref{v92'}) is $\dl_{k-1}$-closed in accordance
with the relation (\ref{v93}). It follows that its summand
$h_{r_k}$ is also $\dl_{k-1}$-closed and, consequently,
$\dl_{k-2}$-closed. Hence it is $\dl_{k-1}$-exact by virtue of the
homology regularity condition. Therefore, $\Delta_{r_k}$ is
reduced only to the summand $G_{r_k}$ linear in antifields. It
follows that the Lagrangian $L_1$ (\ref{w8}) is linear in
antifields. In this case, we have $u^A=\op\dl^\lto{}^A(\cL_e)$,
$u^{r_k}=\op\dl^\lto{}^{r_k}(\cL_e)$ for all indices $A$ and $r_k$
and, consequently,
\be
\up^A=\op\dl^\lto{}^A(\cL_E), \qquad
\up^{r_k}=\op\dl^\lto{}^{r_k}(\cL_E),
\ee
i.e., $\up^0=\up$ is the graded derivation (\ref{w37}) defined by
the Lagrangian (\ref{w133}). Then the nilpotency condition
$\up^0(\up^0)=0$ takes the form
\be\up^0(\op\dl^\lto{}^A(\cL_E))=0, \qquad
\up^0(\op\dl^\lto{}^{r_k}(\cL_E))=0.
\ee
Hence, we obtain
\be
\up^0(L_E)=u(L)+\up^0[\op\dl^\lto{}^A(\cL_E)\ol s_A +
\op\sum_{0\leq k\leq N} \op\dl^\lto{}^{r_k}(\cL_E) \ol c_{r_k}]
d^nx=d_H\si,
\ee
i.e., $\up^0=u_E$ is a variational supersymmetry of the extended
Lagrangian (\ref{w133}).  Thus, this Lagrangian satisfies the
master equation in accordance with Theorem \ref{w39}.
\end{proof}

\bigskip
\bigskip

\noindent {\bf VI. THE GAUGE-FIXING LAGRANGIAN}
\bigskip

Let us further restrict our consideration to a Lagrangian system
which satisfies the condition of Theorem \ref{w130}, i.e., its
gauge and higher-stage gauge supersymmetries form an algebra. Its
total gauge operator
\be
&& u_e=u + \op\sum_{1\leq k\leq N} u_{(k)}, \\
&& u= u^A\frac{\dr}{\dr s^A}, \qquad u^A =\op\sum_{0\leq|\La|}
c^r_\La\eta(\Delta^A_r)^\La, \\
&& u_{(k)}= u^{r_{k-1}}\frac{\dr}{\dr c^{r_{k-1}}}, \qquad
u^{r_{k-1}}=\op\sum_{0\leq|\La|}
c^{r_k}_\La\eta(\Delta^{r_{k-1}}_{r_k})^\La, \qquad k=1,\ldots,N,
\ee
admits a nilpotent extension
\mar{z6}\beq
u_E=u_e+ \xi= u^A\dr_A + \op\sum_{1\leq k\leq N}(u^{r_{k-1}}
+\xi^{r_{k-1}})\dr_{r_{k-1}} + \xi^{r_N}\dr_{r_N}, \label{z6}
\eeq
called the BRST operator. Accordingly, an original Lagrangian $L$
is extended to the Lagrangian
\mar{z7}\beq
L_E= L +[u^A\ol s_A  + \op\sum_{1\leq k\leq N}(u^{r_{k-1}}
+\xi^{r_{k-1}})\ol c_{r_{k-1}} + \xi^{r_N}\ol c_{r_N}]d^nx,
\label{z7}
\eeq
which differs from the Lagrangian (\ref{w133}) in a $d_H$-exact
term. The extended Lagrangian (\ref{z7}) obeys the master equation
\be
\frac{\op\dl^\lto \cL_E}{\dl \ol s_A}\frac{\dl \cL_E}{\dl s^A}d^nx
+ \op\sum_{0\leq k\leq N} \frac{\op\dl^\lto \cL_E}{\dl \ol
c_{r_k}}\frac{\dl \cL_E}{\dl c^{r_k}}d^nx =d_H\si
\ee
and the relation
\be
u_E= \frac{\op\dl^\lto \cL_E}{\dl \ol s_A}\frac{\dr}{\dr s^A} +
\op\sum_{0\leq k\leq N} \frac{\op\dl^\lto \cL_E}{\dl \ol
c_{r_k}}\frac{\dr}{\dr c^{r_k}}.
\ee

The Noether identities (\ref{z20}) show that the extended
Lagrangian $L_E$ (\ref{z7}) is degenerate. Following the BV
quantization procedure, we aim to replace the antifields in $L_E$
with gauge-fixing terms in order to obtain a non-degenerate
Lagrangian \cite{bat,gom}.

For this purpose, let us consider an odd graded density $\Psi
d^nx$ of antifield number 1 which depends on original fields $s^A$
and ghosts $c^{r_k}$, $k=0,\ldots, N$, but not antifields $\ol
s_A$, $\ol c_{r_k}$, $k=0,\ldots, N$. In order to satisfy these
conditions, new field variables must be introduced because all the
ghosts are of negative antifield numbers. Therefore, let us
enlarge the BGDA $P^*\{N\}$ (\ref{w6'}) to the BGDA
\be
\ol P^*\{N\}=\cP^*[\ol E^*_N\cdots\ol E^* \ol V^*_N\cdots\ol V^*
V_N\cdots V\ol Y^*;Q;Y;\ol Q^*E\cdots E_N \ol E^*\cdots\ol E^*_N
\ol V^*\cdots\ol V_N^*],
\ee
possessing the basis
\be
 \{s^A, c^r, c^{r_1}, \ldots,
c^{r_N},c^*_r, c^*_{r_1}, \ldots, c^*_{r_N}, \ol s_A,\ol c_r, \ol
c_{r_1}, \ldots, \ol c_{r_N}\},
\ee
where $[c^*_{r_k}]=[c^{r_k}]$ and Ant$[c^*_{r_k}]=k+1$,
$k=0,\ldots,N$. We agree to call $c^*_r, c^*_{r_1}, \ldots,
c^*_{r_N}$ the antighosts. Then, we choose $\Psi d^nx$ as an
element of $\ol P^{0,n}\{N\}$. It is traditionally called the
gauge-fixing fermion.

Let us replace all the antifields in the Lagrangian $L_E$
(\ref{z7}) with the gauge fixing terms
\be
\ol s_A=\frac{\dl\Psi}{\dl s^A}, \qquad \ol
c_{r_k}=\frac{\dl\Psi}{\dl c^{r_k}}, \qquad k=0,\ldots, N.
\ee
We obtain the Lagrangian
\mar{z13}\beq
L_\Psi= L + [u^A_E \frac{\dl \Psi}{\dl s^A} + \op\sum_{0\leq k\leq
N} u^{r_k}_E\frac{\dl \Psi}{\dl c^{r_k}}]d^nx =L + u_E(\Psi)d^nx
+d_H\si, \label{z13}
\eeq
which is an element of the BGDA
\mar{z18}\beq
\gP^*\{N\}=\cP^*[\ol V^*_N\cdots\ol V^* V_N\cdots V;Q;Y;E\cdots
E_N \ol E^*\cdots\ol E^*_N]\subset \ol P^*\{N\}, \label{z18}
\eeq
possessing the basis
\mar{z11}\beq
\{s^A, c^r, c^{r_1}, \ldots, c^{r_N},c^*_r, c^*_{r_1}, \ldots,
c^*_{r_N}\}.\label{z11}
\eeq
The BRST operator $u_E$ (\ref{z6}) is obviously a graded
derivation of the BGDA $\gP^*\{N\}$. A glance at the equalities
\be
u_E(L_\Psi)=u(L) + u_E(u_E(\Psi)d^nx +d_H\si)d^nx =d_H\si'
\ee
shows that $u_E$ is a variational supersymmetry of the Lagrangian
$L_\Psi$ (\ref{z13}). It however is not a gauge supersymmetry of
$L_\Psi$ if $L_\Psi$ depends on all the ghosts $c^{r_k}$,
$k=0,\ldots,N$, i.e., no ghost is a gauge parameter. Therefore, we
require that
\mar{z14}\beq
\frac{\dl \Psi}{\dl s^A}\neq 0, \qquad \frac{\dl \Psi}{\dl
c^{r_k}}\neq 0, \qquad k=0,\ldots, N-1. \label{z14}
\eeq
In this case, Noether identities for the Lagrangian $L_\Psi$
(\ref{z13}) come neither from the BRST symmetry $u_E$ nor the
equalities (\ref{z20}). One also put
\mar{z15}\beq
\Psi= \op\sum_{0\leq k\leq N} \Psi^{r_k}c^*_{r_k}. \label{z15}
\eeq
Finally, let $h^{r_k r'_k}$ be a non-degenerate bilinear form
$h^{r_k r'_k}$ for each $k=0,\ldots, N-1$ whose coefficients are
either real numbers or functions on $X$. Then a desired
gauge-fixing Lagrangian is written in the form
\mar{z16}\ben
&& L_{GF}=L_\Psi + \op\sum_{0\leq k\leq N}\frac{h_{r_kr'_k}}{2}
\Psi^{r_k}\Psi^{r'_k}d^nx = \label{z16}\\
&& \qquad L + [u^A_E \frac{\dl \Psi}{\dl s^A} + \op\sum_{0\leq
k\leq N} u^{r_k}_E\frac{\dl \Psi}{\dl c^{r_k}}]d^nx +
\op\sum_{0\leq k\leq N}\frac{h_{r_kr'_k}}{2}
\Psi^{r_k}\Psi^{r'_k}d^nx= \nonumber\\
&& \qquad L+ \op\sum_{0\leq k\leq N} u_E(\Psi^{r_k})c^*_{r_k}d^nx
+ \op\sum_{0\leq k\leq N}\frac{h_{r_kr'_k}}{2}
\Psi^{r_k}\Psi^{r'_k}d^nx +d_H\si. \nonumber
\een

The BRST operator $u_E$ (\ref{z6}) fails to be a variational
symmetry of the Lagrangian (\ref{z16}). However, it can be
extended to the graded derivation
\mar{z17}\beq
\wh u=u_E- \op\sum_{0\leq k\leq N}\frac{\op\dr^\lto}{\dr
c^*_{r_k}}h_{r_kr'_k}\Psi^{r'_k} \label{z17}
\eeq
of the BGDA $\gP^*\{N\}$ (\ref{z18}) which is easily proved to be
a variational supersymmetry of the gauge-fixing Lagrangian
(\ref{z16}). We agree to call $\wh u$ (\ref{z17}) the gauge-fixing
BRST symmetry though it is not nilpotent.

Of course, the Lagrangian $L_{GF}$ essentially depends on a choice
of the gauge-fixing fermion $\Psi$ which must satisfy the
conditions (\ref{z14}) and (\ref{z15}). These conditions need not
guarantee that that the Lagrangian $L_{GF}$ is non-degenerate, but
we assume that this is well.

\bigskip
\bigskip

\noindent {\bf VII. QUANTIZATION}
\bigskip

Let us quantize a non-degenerate Lagrangian system
$(\gP^*\{N\},L_{GF})$. Though our results lie in the framework of
perturbed QFT, we start with algebraic QFT.

In algebraic QFT, a quantum field system is characterized by a
topological $^*$-algebra $A$ and a continuous positive form $f$ on
$A$ \cite{borch,hor}. For the sake of simplicity, let us consider
even scalar fields on the Minkowski space $X=\Bbb R^n$. One
associates to them the Borchers algebra $A_\Phi$ of tensor
products of the Schwartz space $\Phi=S(\Bbb R^n)$ of smooth
complex functions of rapid decreasing at infinity on $\Bbb R^n$.
These are complex smooth functions $f$ such that the quantities
\mar{spr453}\beq
|\f|_{k,m}=\op\max_{|\al|\leq k} \op\sup_x(1+x^2)^m|
\frac{\dr^{|\al|}\f}{\dr^{\al_1} x^1\cdots\dr^{\al_n}x^n}|, \qquad
|\al|=\al_1+\cdots +\al_n,  \label{spr453}
\eeq
are finite for all $k,m\in \Bbb N$ for $n$-tuples of natural
numbers $\al=(\al_1,\ldots,\al_n)$. The space $S(\Bbb R^n)$ is
nuclear with respect to the topology determined by the seminorms
(\ref{spr453}). Its topological dual is the space $S'(\Bbb R^n)$
of tempered distributions \cite{piet,bog}. The corresponding
contraction form is written as
\be
\lng \f,\psi\rng=\op\int \f(x) \f(x) d^nx, \qquad \f\in S(\Bbb
R^n), \qquad \psi\in S'(\Bbb R^n).
\ee
The space $S(\Bbb R^n)$ is provided with the non-degenerate
separately continuous Hermitian form
\be
\lng \f|\f'\rng=\int \f(x)\ol\f'(x)d^nx.
\ee
The completion of $S(\Bbb R^n)$ with respect to this form is the
space $L^2_C(\Bbb R^n)$ of square integrable complex functions on
$\Bbb R^n$. We have the rigged Hilbert space
\be
S(\Bbb R^n)\subset L^2_C(\Bbb R^n) \subset S'(\Bbb R^n).
\ee

Let $\Bbb R_n$ denote the dual of $\Bbb R^n$ coordinated by
$(p_\la)$. The Fourier transform
\mar{spr460,1}\ben
&& \f^F(p)=\int \f(x)e^{ipx}d^nx, \qquad px=p_\la x^\la,
\label{spr460}\\
&& \f(x)=\int \f^F(p)e^{-ipx}d_np, \qquad d_np=(2\pi)^{-n}d^np,
\label{spr461}
\een
provides an isomorphism between the spaces $S(\Bbb R^n)$ and
$S(\Bbb R_n)$. The Fourier transform of distributions is defined
by the condition
\be
\int \psi(x)\f(x)d^nx=\int \psi^F(p)\f^F(-p)d_np,
\ee
and is written in the form (\ref{spr460}) -- (\ref{spr461}). It
provides an isomorphism between the spaces of distributions
$S'(\Bbb R^n)$ and $S'(\Bbb R_n)$.

Since $\op\ot^nS(\Bbb R^n)$ is dense in $S(\Bbb R^{nk})$, a state
$f$ of the Borchers algebra $A_\Phi$ is represented by
distributions
\be
f_k(\phi_1\cdots\phi_k)=\int
W_k(x_1,\ldots,x_k)\phi_1(x_1)\cdots\phi_k(x_k) d^nx_1\ldots
d^nx_k, \qquad W_k\in S'(\Bbb R^{nk}).
\ee
In particular, the $k$-point Wightman functions $W_k$ describe
free fields in the Minkowski space. The complete Green functions
characterize quantum fields created at some instant and
annihilated at another one. They are given by the chronological
functionals
\mar{1260}\ben
&& f^c(\phi_1\cdots\phi_k)=\int
W^c_k(x_1,\ldots,x_k)\phi_1(x_1)\cdots\phi_k(x_k)
d^nx_1\ldots d^nx_k, \label{1260}\\
&& W^c_k(x_1,\ldots,x_k)= \op\sum_{(i_1\ldots
i_k)}\th(x^0_{i_1}-x^0_{i_2})
\cdots\th(x^0_{i_{k-1}}-x^0_{i_k})W_k(x_1,\ldots,x_k), \quad
W_{nk}\in S'(\Bbb R^{nk}),\nonumber
\een
where $\th$ is the step function, and the sum runs through all
permutations $(i_1\ldots i_k)$ of the numbers $1,\ldots,k$.
However, the chronological functionals (\ref{1260}) need not be
continuous and positive. At the same time, they issue from the
Wick rotation of Euclidean states of the Borchers algebra $A_\Phi$
describing quantum fields in an interaction zone
\cite{sard91,sard02}. Since the chronological functionals
(\ref{1260}) are symmetric, these Euclidean states are states of
the corresponding commutative tensor algebra $B_\Phi$. This is the
enveloping algebra of the Lie algebra of the group $T(\Phi)$ of
translations in $\Phi$. Therefore one can obtain a state of
$B_\Phi$ as a vector form of a strong-continuous unitary cyclic
representation of $T(\Phi)$ \cite{gel}. Such a representation is
characterized by a positive-definite continuous generating
function $Z$ on $\Phi$. By virtue of the Bochner theorem
\cite{gel}, this function is the Fourier transform
\mar{031}\beq
Z(\phi)=\op\int_{\Phi'}\exp[i \langle\phi,w\rangle]d\mu(w)
\label{031}
\eeq
of a positive measure $\mu$ of total mass 1 on the topological
dual $\Phi'$ of $\Phi$. If the function $\alpha\to Z(\alpha\phi)$
on $\Bbb R$ is analytic at 0 for each $\phi\in \Phi$, a state $F$
of $B_\Phi$ is given by the expression
\mar{w0}\beq
F_k(\phi_1\cdots\phi_k)=i^{-k}\frac{\dr}{\dr \al^1}
\cdots\frac{\dr}{\dr\alpha^k}Z(\alpha^i\phi_i)|_{\alpha^i=0}=\int\langle
\phi_1,w\rangle\cdots\langle \phi_k,w \rangle d\mu(w). \label{w0}
\eeq
Then one can regard $Z$ (\ref{031}) as a generating functional of
Euclidean Green functions $F_k$ (\ref{w0}). A problem is that, if
a field Lagrangian is a polynomial of degree more than two, a
generating functional $Z$ (\ref{031}), Green functions and,
consequently, Ward identities fail to be written in an explicit
form.

Therefore, let us quantize the above mentioned non-degenerate
Lagrangian system $(\gP^*\{N\},L_{GF})$ in the framework of
perturbed QFT. We assume that $L_{GF}$ is a Lagrangian of
Euclidean fields on $X=\Bbb R^n$, coordinated by $(x^\la)$. The
key point is that an algebra of Euclidean quantum fields is graded
commutative, and there are homomorphisms of the graded commutative
ring $\gP^0\{N\}$ of classical fields to this algebra. These
monomorphisms enable one to define the BRST operator $\wh u$
(\ref{z17}) on Euclidean quantum fields.

Let $\cQ$ be the graded complex vector space whose basis is the
basis (\ref{z11}) for the BGDA $\gP^*\{N\}$. The common symbols
$q^a$ further stand for elements of this basis. Let us consider
the tensor product
\mar{z40}\beq
\Phi=\cQ\ot S'(\Bbb R^n) \label{z40}
\eeq
of the graded vector space $\cQ$ and the space $S'(\Bbb R^n)$ of
distributions on $\Bbb R^n$. One can think of elements $\Phi$
(\ref{z40}) as being $\cQ$-valued distributions on $\Bbb R^n$. Let
us consider the subspace $T(\Bbb R^n)\subset S'(\Bbb R^n)$ of
functions $\exp\{ipx'\}$, $p\in \Bbb R_n$, which are generalized
eigenvectors of translations in $\Bbb R^n$ acting on $S(\Bbb
R^n)$. Let us denote $f^a_p=q^a\ot\exp\{ipx'\}$. Then any element
$\f$ of $\Phi$ can be written in the form
\mar{z42}\beq
\f(x')=q^a\ot\f_a(x')=\int \f_a(p)\f_p^a d_np, \label{z42}
\eeq
where $\f_a(p)\in S'(\Bbb R_n)$ are the Fourier transforms of
$\f_a(-x')$. For instance, we have the $\cQ$-valued distributions
\mar{z32,43}\ben
&& \f^a_x(x')=\int \f^a_p e^{-ipx}d_np=q^a\ot \dl(x-x'),
\label{z32}\\
&& \f^a_{x\La}(x')=\int (-i)^k p_{\la_1}\cdots p_{\la_k}\f^a_p
e^{-ipx}d_np. \label{z43}
\een

In the framework of perturbed Euclidean QFT, we associate to a
non-degenerate Lagrangian system $(\gP^*\{N\},L_{GF})$ the graded
commutative tensor algebra $B_\Phi$ generated by elements of the
graded vector space $\Phi$ (\ref{z40}) and the following state
$\lng.\rng$ of $B_\Phi$. For any $x\in X$, there is a homomorphism
\mar{z45}\beq
\g_x: f_{a_1\ldots a_r}^{\La_1\ldots\La_r} q^{a_1}_{\La_1}\cdots
q^{a_r}_{\La_r} \mapsto f_{a_1\ldots a_r}^{\La_1\ldots\La_r}(x)
\f^{a_1}_{x\La_1}\cdots \f_{x\La_r}^{a_r}, \qquad f_{a_1\ldots
a_r}^{\La_1\ldots\La_r}\in C^\infty(X), \label{z45}
\eeq
of the $C^\infty(X)$-ring $\gP^0\{N\}$ to the algebra $B_\Phi$
which sends elements $q^a_\La\in \gP^0\{N\}$ to the elements
$\f^a_{x\La}\in B_\Phi$ (\ref{z43}), and replaces coefficient
functions $f$ of elements of $\gP^0\{N\}$ with their values $f(x)$
at a point $x$. It should be emphasized that $f_{a_1\ldots
a_r}^{\La_1\ldots\La_r}(x) \f^{a_1}_{x\La_1}\cdots
\f_{x\La_r}^{a_r}$ in the expression (\ref{z45}) is the graded
commutative tensor product of distributions, but not their product
which is ill defined. Then the above mentioned state $\lng.\rng$
of $B_\Phi$ is given by symbolic functional integrals
\mar{z31,',47}\ben
&& \lng \f_1\cdots \f_k\rng=\frac{1}{\cN}\int \f_1\cdots \f_k
\exp\{-\int \cL_{GF}(\f^a_p)d^nx\}\op\prod_p
[d\phi_p^a], \label{z31}\\
&& \cN=\int \exp\{-\int \cL_{GF}(\f^a_p)d^nx\}\op\prod_p
[d\phi_p^a], \label{z31'}\\
&&
\cL_{GF}(\f^a_p)=\cL_{GF}(\f^a_{x\La})=\cL_{GF}(x,\g_x(q^a_\La)),
\label{z47}
\een
where $\f_i$ and $\g_x(q^a_\La)=\f^a_{x\La}$ are given by the
formulas (\ref{z42}) and (\ref{z43}), respectively. The forms
(\ref{z31}) are expressed into the forms
\mar{z48,49}\ben
&& \lng\f^{a_1}_{p_1}\cdots \f^{a_k}_{p_k}\rng=\frac{1}{\cN}\int
\f^{a_1}_{p_1}\cdots \f^{a_k}_{p_k} \exp\{-\int
\cL_{GF}(\f^a_p)d^nx\}\op\prod_p [d\phi_p^a], \label{z48}\\
&& \lng\f^{a_1}_{x_1}\cdots \f^{a_k}_{x_k}\rng=\frac{1}{\cN}\int
\f^{a_1}_{x_1}\cdots \f^{a_k}_{x_k} \exp\{-\int
\cL_{GF}(\f^a_{x\La})d^nx\}\op\prod_x [d\phi_x^a], \label{z49}
\een
which provide Euclidean Green functions. It should be emphasized
that, in contrast with a measure $\m$ in the expression
(\ref{031}), the term $\op\prod_p [d\phi_p^a]$ in the formulas
(\ref{z31}) -- (\ref{z31'}) fail to be a true measure on $T(\Bbb
R^n)$ because the Lebesgue measure on infinite-dimensional vector
spaces need not exist. Nevertheless, treated like integrals over a
finite-dimensional vector space, the functional integrals
(\ref{z48})  - (\ref{z49}) restart Euclidean Green functions in
the Feynman diagram technique. Certainly, these Green functions
are singular, unless regularization and renormalization techniques
are involved.

\bigskip
\bigskip

\noindent {\bf VIII. WARD IDENTITIES}
\bigskip

Since the graded derivation (\ref{z17})
\be
\wh u=\op\sum_{0\leq|\La|}\wh u^a_\La(q^b_\Si)\dr_a^\La
\ee
of the ring $\gP^0\{N\}$ is a $C^\infty(X)$-linear morphism over
$\id X$, it induces the graded derivation
\mar{z50}\beq
\wh u_x= \g_x\circ \wh u\circ \g^{-1}_x: \f^a_{x\La}\to (x,
q^a_\La))\to \wh u^a_\La(x,q^b_\Si)\to \wh
u^a_\La(x,\g_x(q^b_\Si))=\wh u^a_{x\La}(\f^b_{x\Si}) \label{z50}
\eeq
of the range $\g_x(\gP^0\{N\})\subset B_\Phi$ of the homomorphism
$\g_x$ (\ref{z45}) for each $x\in X=\Bbb R^n$. The maps $\wh u_x$
(\ref{z50}) yield the maps
\be
&& \wh u_p: \f^a_p=\int \f^a_x
e^{ipx}d^nx \to \int \wh u_x(\f^a_x)e^{ipx}d^nx= \int \wh
u^a_x(\f^b_{x\Si})e^{ipx}d^nx = \\
&& \qquad  \int \wh u_x^a(\int (-i)^kp'_{\si_1}\cdots
p'_{\si_k}\f^b_{p'}e^{-ip'x}d_np') e^{ipx}d^nx= \wh u^a_p, \qquad
p\in\Bbb R_n,
\ee
and, as a consequence, the graded derivation
\be
\wh u(\f)=\int\f_a(p)\wh u(\f^a_p) d_np= \int\f_a(p)\wh u^a_pd_np
\ee
of the algebra $B_\Phi$. We agree to call it quantum BRST
transformation. It can be written in the symbolic form
\mar{z53,4}\ben
&& \wh u= \int u^a_p\frac{\dr}{\dr \f^a_p}d_np, \qquad \frac{\dr
\f^b_{p'}}{\dr \f^a_p} =\dl^b_a\dl(p'-p), \label{z53}\\
&& \wh u= \int u^a_x\frac{\dr}{\dr \f^a_x}d^nx, \qquad \frac{\dr
\f^b_{x'\La}}{\dr \f^a_x}= \dl^b_a\frac{\dr}{\dr x'^{\la_1}}
\cdots \frac{\dr}{\dr x'^{\la_k}} \dl(x'-x). \label{z54}
\een

Let $\al$ be an odd element. We consider the automorphism
\be
\wh U=\exp\{\al \wh u\}=\id +\al\wh u
\ee
of the algebra $B_\Phi$. This automorphism yields a new state
$\lng.\rng'$ of $B_\Phi$ given by the relations
\be
&& \lng \f_1\cdots \f_k\rng= \lng \wh U(\f_1)\cdots \wh
U(\f_k)\rng'= \\
&& \qquad \frac{1}{\cN'}\int \wh U(\f_1)\cdots \wh U(\f_k)
\exp\{-\int \cL_{GF}(\wh U(\f^a_p))d^nx\}\op\prod_p [d\wh
U(\phi_p^a)], \\
&& \cN'=\int \exp\{-\int \cL_{GF}(\wh U(\f^a_p))d^nx\}\op\prod_p
[d\wh U(\phi_p^a)].
\ee
Let us apply these relations to the Green functions (\ref{z48}) --
(\ref{z49}).

Since the graded derivation $\wh u$ (\ref{z17}) is a variational
supersymmetry of the Lagrangian $L_{GF}$ (\ref{z18}), we obtain
from the relations (\ref{z47}) that
\be
&& \int \cL_{GF}(\wh U(\f^a_{x\La}))d^nx =\int
\cL_{GF}(\f^a_{x\La})d^nx, \\
&& \int \cL_{GF}(\wh U(\f^a_p))d^nx =\int \cL_{GF}(\f^a_p)d^nx.
\ee
It is a property of symbolic functional integrals that
\be
&& \op\prod_p[d\wh U(\phi_p^a)]=(1+\al\int \frac{\dr \wh
u^a_p}{\dr \f^a_p}d_np)\op\prod_p[d\phi_p^a]=(1+\al {\rm Sp}(\wh
u))
\op\prod_p[d\phi_p^a], \\
&& \op\prod_x[d\wh U(\phi_x^a)]=(1+\al\int \frac{\dr \wh
u^a_x}{\dr \f^a_x}d^nx)\op\prod_x[d\phi_x^a]=(1+\al {\rm Sp}(\wh
u)) \op\prod_x[d\phi_x^a].
\ee
Then the desired Ward identities for the Green functions
(\ref{z48}) -- (\ref{z49}) read
\mar{z61,2}\ben
&& \lng\wh u(\f^{a_1}_{p_1}\cdots \f^{a_k}_{p_k})\rng
+\lng\f^{a_1}_{p_1}\cdots \f^{a_k}_{p_k}{\rm Sp}(\wh u)\rng- \lng
\f^{a_1}_{p_1}\cdots \f^{a_k}_{p_k}\rng\lng{\rm Sp}(\wh u)\rng =0,
\label{z61}\\
&& \op\sum_{i=1}^k(-1)^{[a_1]+\cdots +[a_{i-1}]}
\lng\f^{a_1}_{p_1}\cdots \f^{a_{i-1}}_{p_{i-1}}\wh u^{a_i}_{p_i}
\f^{a_{i+1}}_{p_{i+1}}\cdots \f^{a_k}_{p_k}\rng +\nonumber\\
&& \qquad \lng\f^{a_1}_{p_1}\cdots \f^{a_k}_{p_k}\int \frac{\dr
\wh u^a_p}{\dr \f^a_p}d_np\rng - \lng\f^{a_1}_{p_1}\cdots
\f^{a_k}_{p_k}\rng\lng\int \frac{\dr \wh u^a_p}{\dr
\f^a_p}d_np\rng=0,
\nonumber\\
&& \lng\wh u(\f^{a_1}_{x_1}\cdots \f^{a_k}_{x_k})\rng
+\lng\f^{a_1}_{x_1}\cdots \f^{a_k}_{x_k}{\rm Sp}(\wh u)\rng- \lng
\f^{a_1}_{x_1}\cdots \f^{a_k}_{x_k}\rng\lng{\rm Sp}(\wh u)\rng =0,
\label{z62}\\
&& \op\sum_{i=1}^k(-1)^{[a_1]+\cdots +[a_{i-1}]}
\lng\f^{a_1}_{x_1}\cdots \f^{a_{i-1}}_{x_{i-1}}\wh u^{a_i}_{x_i}
\f^{a_{i+1}}_{x_{i+1}}\cdots \f^{a_k}_{x_k}\rng +\nonumber\\
&&\qquad \lng\f^{a_1}_{x_1}\cdots \f^{a_k}_{x_k}\int \frac{\dr \wh
u^a_x}{\dr \f^a_x}d^nx\rng - \lng\f^{a_1}_{x_1}\cdots
\f^{a_k}_{x_k}\rng\lng\int \frac{\dr \wh u^a_x}{\dr
\f^a_x}d^nx\rng=0. \nonumber
\een

A glance at the expressions (\ref{z61}) -- (\ref{z62}) shows that
the Ward identities contain the anomaly, in general, because the
measure terms of symbolic functional integrals need not be BRST
invariant. If Sp$(\wh u)$ is either a finite or infinite number,
the Ward identities
\mar{z63,4}\ben
&& \lng\wh u(\f^{a_1}_{p_1}\cdots \f^{a_k}_{p_k})\rng =
\op\sum_{i=1}^k(-1)^{[a_1]+\cdots +[a_{i-1}]}
\lng\f^{a_1}_{p_1}\cdots \f^{a_{i-1}}_{p_{i-1}}\wh u^{a_i}_{p_i}
\f^{a_{i+1}}_{p_{i+1}}\cdots \f^{a_k}_{p_k}\rng=0, \label{z63}\\
&& \lng\wh u(\f^{a_1}_{x_1}\cdots \f^{a_k}_{x_k})\rng=
\op\sum_{i=1}^k(-1)^{[a_1]+\cdots +[a_{i-1}]}
\lng\f^{a_1}_{x_1}\cdots \f^{a_{i-1}}_{x_{i-1}}\wh u^{a_i}_{x_i}
\f^{a_{i+1}}_{x_{i+1}}\cdots \f^{a_k}_{x_k}\rng =0 \label{z64}
\een
are free of this anomaly.

Clearly, all the expressions (\ref{z61}) -- (\ref{z64}) are
singular, unless regularization and renormalization procedures are
involved. In the present work, we concern neither these procedures
nor the Wick rotation of the Ward identities (\ref{z61}) --
(\ref{z62}).

\bigskip
\bigskip

\noindent {\bf XI. EXAMPLE. SUPERSYMMETRIC YANG--MILLS THEORY}
\bigskip

Let $\cG=\cG_0\oplus \cG_1$ be a finite-dimensional real Lie
superalgebra with a basis $\{e_r\}$, $r=1,\ldots,m,$ and real
structure constants $c^r_{ij}$. For the sake of simplicity, the
Grassmann parity of $e_r$ is denoted $[r]$. Recall the standard
relations
\be
&& c^r_{ij}=-(-1)^{[i][j]}c^r_{ji}, \qquad [r]=[i]+[j],\\
&& (-1)^{[i][b]}c^r_{ij}c^j_{ab} + (-1)^{[a][i]}c^r_{aj}c^j_{bi} +
(-1)^{[b][a]}c^r_{bj}c^j_{ia}=0.
\ee
Let us also introduce the modified structure constants
\mar{z90}\beq
\ol c^r_{ij}=(-1)^{[i]}c^r_{ij}, \qquad \ol
c^r_{ij}=(-1)^{([i]+1)([j]+1)}\ol c^r_{ji}. \label{z90}
\eeq
Given the universal enveloping algebra $\ol \cG$ of $\cG$, we
assume that there is an invariant even quadratic element
$h^{ij}e_ie_j$ of $\ol\cG$ such that the matrix $h^{ij}$ is
non-degenerate. All Lagrangians are further considered up to
$d_H$-exact terms.

The Yang--Mills theory of gauge potentials on $X=\Bbb R^n$
associated to the Lie superalgebra $\cG$ is described by the BGDA
$\cP^*[Q,Y]$ where
\be
Q=(X\times \cG_1)\op\ot_X T^*X, \qquad Y= (X\times \cG_0)\op\ot_X
T^*X.
\ee
Its basis is $\{a^r_\la\}$, $[a^r_\la]=[r]$. There is the
canonical decomposition of the first jets of its elements
\be
a^r_{\la\m}=\frac12(\cF^r_{\la\m} +
\cS^r_{\la\m})=\frac12(a^r_{\la\m}-a^r_{\m\la} +c^r_{ij}a^i_\la
a^j_\m) +\frac12(a^r_{\la\m}+ a^r_{\m\la} -c^r_{ij}a^i_\la
a^j_\m).
\ee
Then the Euclidean Yang--Mills Lagrangian takes the form
\be
L_{YM}=\frac14
h_{ij}\eta^{\la\m}\eta^{\bt\nu}\cF^i_{\la\bt}\cF^j_{\m\nu}d^nx,
\ee
where $\eta$ is the Euclidean metric on $\Bbb R^n$. Its
variational derivatives $\cE_r^\la$ obey the irreducible Noether
identities
\be
 -c^r_{ji}a^i_\la\cE_r^\la - d_\la\cE_j^\la=0.
\ee
Therefore, we enlarge the BGDA $\cP^*[Q,Y]$ to the BGDA
\mar{z73}\beq
P^*\{0\}=\cP^*[\ol E^* V\ol Y^*;Q;Y;\ol Q^*E\ol V^*], \qquad
V=X\times \cG_1, \qquad E=X\times \cG_0, \label{z73}
\eeq
whose basis
\be
\{a^r_\la, c^r, \ol a^\la_r, \ol c_r\}, \qquad [c^r]=([r]+1){\rm
mod}\,2, \qquad [\ol a^\la_r]=[\ol c_r]=[r],
\ee
consists of gauge potentials $a^r_\la$, ghosts $c^r$ of ghost
number 1, and antifields $\ol a^\la_r$, $\ol c_r$ of antifield
numbers 1 and 2, respectively. Then, the Noether operators
$\Delta_r$ (\ref{v63}), the total gauge operator $u_e$
(\ref{w108}), and the Lagrangian $L_e$ (\ref{w8}) read
\be
&& \Delta_j= -c^r_{ji}a^i_\la\ol a_r^\la - \ol a_{\la j}^\la,\\
&& u_e=u^r_\la\dr^\la_r= (-c^r_{ji}c^ja^i_\la + c^r_\la)\dr^\la_r,\\
&& L_e= L_{YM} - c^j(c^r_{ji}a^i_\la\ol a_r^\la + \ol a_{\la
j}^\la)d^nx= L_{YM} + (-c^r_{ji}c^ja^i_\la + c^r_\la)\ol
a^\la_rd^nx +d_H\si.
\ee

The total gauge operator $u_e$ admits the nilpotent extension
\be
u_E=u_e +\xi= (-c^r_{ji}c^ja^i_\la + c^r_\la)\dr^\la_r -\frac12
\ol c^r_{ij}c^ic^j\dr_r,
\ee
where $\ol c^r_{ij}$ are the modified structure constants
(\ref{z90}). Then, the extended Lagrangian
\mar{z78}\beq
L_E=L_{YM}+ (-c^r_{ij}c^ja^i_\la + c^r_\la)\ol a^\la_rd^nx
-\frac12 \ol c^r_{ij}c^ic^j\ol c_rd^nx, \label{z78}
\eeq
obeys the master equation.

Passing to the gauge-fixing Lagrangian, we enlarge the BGDA
$P^*[0]$ (\ref{z73}) to the BGDA
\be
\ol P^*\{0\}=\cP^*[\ol E^* \ol V^*V\ol Y^*;Q;Y;\ol Q^*E\ol E^*\ol
V^*],
\ee
possessing the basis
\be
\{a^r_\la, c^r, c^*_r, \ol a^\la_r, \ol c_r\}, \qquad
[c^*_r]=[c^r].
\ee
Let us choose the gauge-fixing fermion
\be
\Psi = \frac12\eta^{\la\m}\cS^r_{\la\m}c^*_r=
\eta^{\la\m}a^r_{\la\m}c^*_r,
\ee
and replace the antifields in the extended Lagrangian $L_E$
(\ref{z78}) with the terms
\be
\ol a^\la_r=\frac{\dl \Psi}{\dl a^r_\la}=-\eta^{\la\m}c^*_{\m r},
\qquad \ol c_r=0.
\ee
We come to the Lagrangian $L_\Psi$ (\ref{z13}) which reads
\be
L_\Psi=L_{YM}-\eta^{\la\m}(-c^r_{ij}c^ja^i_\la + c^r_\la)c^*_{\m
r}d^nx = L_{YM} + (-1)^{[r]+1}\eta^{\la\m}c^*_r
d_\m(-c^r_{ij}c^ja^i_\la + c^r_\la)d^nx+ d_H\si.
\ee
It is brought into the form
\be
L_\Psi=L_{YM} +c^*_r M_j^r c^jd^nx,
\ee
where
\be
M_i^r=(-1)^{[r]+1}\eta^{\la\m}(c^r_{ij}(a^i_{\m\la}+ a^i_\la d_\m)
+ \dl^r_j d_{\m\la})
\ee
is a second order differential operator acting on the ghosts
$c^j$. Finally, we write the gauge-fixing Lagrangian
\be
&& L_{GF}=L_{YM} +c^*_r M_j^r c^jd^nx +\frac18
h_{ij}\eta^{\la\m}\eta^{\bt\nu}\cS^i_{\la\m}\cS^j_{\bt\nu}d^nx=\\
&& \qquad L_{YM} + [(-1)^{[r]+1}\eta^{\la\m}c^*_r
d_\m(-c^r_{ij}c^ja^i_\la + c^r_\la) + \frac12
h_{ij}\eta^{\la\m}\eta^{\bt\nu}a^i_{\la\m}a^j_{\bt\nu}]d^nx,
\ee
which possesses the BRST symmetry
\be
\wh u= (-c^r_{ij}c^ja^i_\la + c^r_\la)\frac{\dr}{\dr a_\la^r}
-\frac12 \ol c^r_{ij}c^ic^j\frac{\dr}{\dr c^r} +(-1)^{[j]}
h_{ij}\eta^{\la\m}a^i_{\la\m}\frac{\dr}{\dr c^*_j}.
\ee

Quantizing this Lagrangian system in the framework of Euclidean
perturbed QFT , we come to the graded commutative tensor algebra
$B_\Phi$ generated by the elements $\{a_{x\la}^r, c_x^r,
c^*_{xr}\}$. Its states $\lng\f\rng$, $\f\in B_\Phi$, are given by
functional integrals
\be
&& \lng \f\rng=\frac{1}{\cN}\int \f \exp\{-\int
\cL_{GF}(a^r_{x\La\la}, c^r_{x\La}, c^*_{xr})d^nx\}\op\prod_x
[da^r_{x\la}][dc^r_x][dc^*_{xr}], \\
&& \cN=\int \exp\{-\int \cL_{GF}(a^r_{x\La\la}, c^r_{x\La},
c^*_{xr})d^nx\}\op\prod_x [da^r_{x\la}][dc^r_x][dc^*_{xr}],\\
&& \cL_{GF}=\cL_{YM} + (-1)^{[r]+1}\eta^{\la\m}c^*_{xr}
d_\m(-c^r_{xij}c^j_xa^i_{x\la} + c^r_{x\la}) + \frac12
h_{ij}\eta^{\la\m}\eta^{\bt\nu}a^i_{x\la\m}a^j_{x\bt\nu}.
\ee
Accordingly, the quantum BRST transformation (\ref{z54}) reads
\be
\wh u= \int[(-c^r_{ij}c^j_xa^i_{x\la} + c^r_{x\la})\frac{\dr}{\dr
a_{x\la}^r} -\frac12 \ol c^r_{ij}c^i_xc^j_x\frac{\dr}{\dr c^r_x}
+(-1)^{[j]} h_{ij}\eta^{\la\m}a^i_{x\la\m}\frac{\dr}{\dr
c^*_{xj}}]d^nx.
\ee
It is readily observed that Sp$(\wh u)=0$. Therefore, we obtain
the Ward identities $\lng\wh u(\f)\rng=0$ (\ref{z64}) without
anomaly.

\end{document}